\documentstyle[prb,aps]{revtex}
\begin{document}
\draft
\title{Fine structure of excitons in a quantum well in the presence of a nonhomogeneous magnetic
field}
\author{J. A. K. Freire\cite{AddJAKF}, F. M.
Peeters\cite{AddFMP}, and A. Matulis\cite{AddMat}}
\address{Departement Natuurkunde, Universiteit Antwerpen (UIA),
Universiteitsplein 1, B-2610 Antwerpen, Belgium}
\author{V. N. Freire and G. A. Farias}
\address{Departamento de F\'{\i}sica, Universidade Federal do Cear\'a, Centro de Ci\^encias
Exatas, Campus do Pici, Caixa Postal 6030, 60455-760 Fortaleza,
Cear\'a, Brazil}
\date{\today}
\maketitle
\begin{abstract}
The trapping of excitons in a semiconductor quantum well due to a
circular symmetric nonhomogeneous magnetic field is studied. The
effect of the spin state of the exciton on its trapping energy is
analyzed, and the importance of the interaction of the orbital and
spin Zeeman effect as compared to the diamagnetic term in the
exciton Hamiltonian is emphasized. Magnetic field profiles are
considered, which can experimentally be created through the
deposition of ferromagnetic disks on top of a semiconductor
heterostructure. This setup gives rise to a magnetic dipole type
of profile in the $xy$ plane of the exciton motion. We find that
the spin direction of the exciton influences its localization by
changing the confinement region in the effective potential. The
exciton confinement increases with magnetic field intensity, and
this is more pronounced when the exciton g-factor is different
from zero. The numerical calculations are performed for
GaAs/Al$_{x}$Ga$_{1-x}$As quantum wells and we show that it open
up a new realistic path for experiments designed to probe exciton
trapping in semiconductors.
\end{abstract}
\section{Introduction}
The trapping and guiding of atoms has been the subject of a number
of experimental and theoretical works during the last few
years.\cite{Wie,Hind} Among the various methods that have been
applied to the trapping of atoms, the magnetic one has been
clearly recognized as the most powerful.\cite{Mew,Zimmer} The
possibility of exciton trapping in semiconductors has been
actively pursued for many years.\cite{Wolfe,Butov} Recently,
nonhomogeneous stress was used in order to create an energy
minimum for the trapping of excitons in GaAs quantum
wells.\cite{Negoit} Several works have argued that the utilization
of  homogeneous magnetic fields during the trapping of excitons in
semiconductors quantum wells will enhance Bose-Einstein effects in
a gas of excitons.\cite{Koro} In particular, Ref. [9] reported a
strong increase of the exciton confinement properties above a
critical threshold of uniform magnetic field. An alternative
approach to localize excitons is by using nonhomogeneous magnetic
fields. In a preceding paper,\cite{JAKF} we discussed the exciton
lateral confinement in a GaAs two-dimensional electron gas (2DEG)
in the presence of nonhomogeneous magnetic fields. We showed that
excitons can be trapped in magnetic field inhomogeneities, with a
confinement degree that depends strongly on their angular
momentum. Experimentally, nonhomogeneous magnetic fields were used
to control the confinement of excitons in quantum wells, where
spatially resolved photoluminescence (PL) and PL-excitation
spectroscopy was utilized to measure the field gradient effect in
the exciton lateral motion. \cite{Fabio}

With the continuing improvements in the experimental realization
of spatially nonhomogeneous magnetic fields, several magnetic
structures which can be used for the trapping of particles have
been proposed,\cite{PeetersRev} such as: magnetic quantum dots
produced by a scanning tunneling microscope lithographic
technique,\cite{McCord} magnetic superlattices created through the
patterning of ferromagnetic materials integrated with
semiconductors,\cite{Krishnan} type-II superconducting materials
deposited on conventional heterostructures,\cite{Bend} and
nonplanar 2DEG systems grown by molecular-beam epitaxy.\cite{Lead}
In particular, a magnetic dipole type of profile, which
experimentally can be created by ferromagnetic materials deposited
on top of semiconductor heterostructures, has attracted a great
deal of interest.\cite{JAKF,Dubonos} This system is essentially
different from the others since the local nonhomogeneous magnetic
field has zero average magnetic field strength. Furthermore, the
magnetized disks can create much stronger magnetic field
inhomogeneities than superconducting one.

In our previous paper, \cite{JAKF} we neglected the spin of the
electron and the heavy-hole and considered only their orbital
motion. We showed that this could already induce a weak
confinement of the excitons for special configurations of the
nonhomogeneous magnetic field profile. In the trapping of atoms,
their spin/total angular momentum is the key quantity which leads
to their confinement in a nonhomogeneous magnetic field.
Therefore, we expect that in the case of excitons, the inclusion
of the spin of the particles will lead to an increased
confinement.

In this work, we report on a detailed analysis of the trapping of
excitons in a GaAs/Al$_{x}$Ga$_{1-x}$As quantum well in a
nonhomogeneous magnetic field taking into account spin effects. A
magnetized disk deposited on top of a quantum well with a
homogeneous magnetic field applied parallel to its growth
direction $z$, i.e., perpendicular to the magnetic disk, generates
the nonhomogeneous magnetic field profile used in the present
calculations. This system gives rise to a magnetic dipole type of
profile in the $xy$ plane of the exciton
motion.\cite{JAKF,Dubonos,GeimPt} The effects of the well
confinement and the importance of the orbital and spin Zeeman
splitting, and the diamagnetic contributions to the exciton
trapping energy and the wave function are analysed.

The paper is organized as follows. In Sec. II the theoretical
procedure used to obtain the exciton Hamiltonian in a quantum well
in nonhomogeneous magnetic fields is given. The circular magnetic
trap used in this work is also presented in Sec. II. The
calculation method to obtain the exciton trapping energy, the
exciton effective mass and its effective confinement potential are
discussed in Sec. III. Our numerical results for
GaAs/Al$_{x}$Ga$_{1-x}$As quantum wells are presented and
discussed in Sec. IV. These theoretical results strongly suggest
that it is possible to realize experimentally exciton trapping
using nonhomogeneous magnetic fields. Our results are summarized
in Sec. V.
\section{Theoretical Model}

\subsection{Exciton Hamiltonian}

The exciton Hamiltonian describing the electron and heavy-hole
motion in a quantum well in nonhomogeneous magnetic fields can be
written as follows:

\begin{equation}
H=H^{2D}({\mathbf r}_{e},{\mathbf r}_{h})+W(r,z_{e,h})+H^{\bot
}(z_{e,h})+H^{m_z }({\mathbf r}_{e},{\mathbf r}_{h}), \label{Ham1}
\end{equation}
where
\begin{mathletters}
\label{alleq2}
\begin{eqnarray}
H^{2D}({\mathbf r}_{e},{\mathbf r}_{h}) &=&\frac{\hbar
^{2}}{2m_{e,\Vert }^{\ast }} \left\{ -i\nabla
_{e}^{2D}+\frac{e}{\hbar c}{\mathbf A}({\mathbf r}_{e})\right\}
^{2}+\frac{\hbar ^{2}}{2m_{h,\Vert }^{\ast }}\left\{ -i\nabla
_{h}^{2D}- \frac{e}{\hbar c}{\mathbf A}({\mathbf r}_{h})\right\}
^{2}-\frac{\gamma e^{2}}{ \varepsilon r}, \\ W(r,z_{e,h})
&=&\frac{\gamma e^{2}}{\varepsilon r}-\frac{e^{2}}{\varepsilon
\sqrt{ r^{2}+(z_{e}-z_{h})^{2}}}, \\ H^{\bot }(z_{e,h})
&=&-\frac{\hbar ^{2}}{2m_{e,\bot }^{\ast }}\frac{\partial ^{2}}{
\partial z_{e}^{2}}+V_{e}(z_{e})-\frac{\hbar ^{2}}{2m_{h,\bot }^{\ast }}
\frac{\partial ^{2}}{\partial z_{h}^{2}}+V_{h}(z_{h}),
\\ H^{m_z }({\mathbf r}_{e},{\mathbf r}_{h}) &=&\mu
_{B}\sum_{i=x,y,z} g_{e,i}S_{e,i}B_{i}{\mathbf (r}_{e}{\mathbf
)}-2\mu _{B}\sum_{i=x,y,z} \left(
k_{i}J_{h,i}+q_{i}J_{h,i}^{3}\right) B_{i}({\mathbf r}
_{h})-\frac{2}{3}\sum_{i=x,y,z} c_{i}S_{e,i}J_{h,i}. \label{eq2}
\end{eqnarray}
\end{mathletters}

The in-plane Hamiltonian [$H^{2D}({\mathbf r}_{e},{\mathbf
r}_{h})$] describes the electron and hole motion in the $xy$ plane
in a nonhomogeneous magnetic field, which is represented by the
vector potential ${\mathbf A}({\mathbf r})$. We will make use of a
variational approach\cite{Bao} to assume that the difference
between the 2D and 3D Coulomb interactions [$W(r,z_{e,h})$] can be
made very small by the choice of an optimum value for $\gamma
$,\cite{Spi} which is a variational parameter that is calculated
from the average of $ W(r,z_{e,h})$ over the exciton wave
function. The perpendicular contribution [$H^{\bot }(z_{e,h})$]
describes the exciton confinement in the quantum well, i.e., in
the $z$ direction, which is not affected by the magnetic field.
The exciton spin interaction with the magnetic field is described
by the spin Hamiltonian [$H^{m_z }({\mathbf r})$]. The terms in
the latter correspond to the electron and heavy hole Zeeman spin
interaction, and to the spin-spin coupling energy,
respectively.\cite{Kest}

In Eq. (\ref{alleq2}), $m_{e,\bot }^{\ast }$, $m_{h,\bot }^{\ast
}$ ($m_{e,\Vert }^{\ast }$, $m_{h,\Vert }^{\ast }$) are the
perpendicular (in-plane) electron and heavy hole
effective masses,$\ S_{e,i}$ and $J_{h,i}$ ($g_{e,i}$ and $%
k_{i},q_{i}$) are related to the electron and heavy hole spin
(Luttinger Zeeman splitting constants), respectively, $m_z
=S_{e,i}+J_{h,i} = \pm 1 , \pm 2 $, is the total spin angular
momentum,$\ \mu _{B}=e\hbar /2m_{e,\Vert }^{\ast }c$ is the Bohr
magneton, and $c_{i}$ is the spin-spin coupling constant related
to the zero field spin interaction. We will assume an isotropic
dispersion for electrons and holes. Notice that ${\mathbf r}_{e}$,
${\mathbf r}_{h}$ are 2D coordinates and give the $xy$ plane
position of the electron and hole, respectively.

To simplify the in-plane [$H^{2D}({\mathbf r}_{e},{\mathbf
r}_{h})$] and spin [$H^{m_z }({\mathbf r}_{e},{\mathbf r}_{h})$]
Hamiltonian, we introduce the exciton relative and center-of mass
coordinates, ${\mathbf r} = {\mathbf r}_{e} - {\mathbf r}_{h}$,
and $\ {\mathbf R} = \left(m_{e}^{\ast}{\mathbf r}_{e} +
m_{h}^{\ast}{\mathbf r}_{h}\right) / M $, respectively, with $ M =
(m_{e}^{\ast}+m_{h}^{\ast})$ the exciton mass. Following Ref.
[10], we apply the adiabatic approach\cite{Adiabat} in which we
assume that the exciton relative motion is fast as compared to the
center-of-mass motion. Following the approach of Freire {\it et
al}.,\cite{JAKF} the $H^{2D}({\mathbf r}_{e},{\mathbf r}_{h}) $
Hamiltonian can be separated in a center-of-mass and a relative
Hamiltonian, i.e., $H^{2D}({\mathbf r}_{e},{\mathbf r}_{h}) =
H^{CM}({\mathbf R})+H_{\gamma }^{r}({\mathbf r },{\mathbf
R},\nabla _{R})$, with $H^{CM}({\mathbf R}) = -(\hbar^{2}/2M)
\nabla_{R}^{2}$, and:

\begin{eqnarray}
H_{\gamma }^{r}({\mathbf r},{\mathbf R},\nabla _{R}) =-\frac{\hbar
^{2}}{2\mu } \nabla _{r}^{2}-\frac{\gamma e^{2}}{\varepsilon r} +
W_1  + W_2, \label{relH}
\end{eqnarray}
where $W_1 = ({e}/{2\mu c})\xi {\mathbf B(R)}\cdot {\mathbf L}
\nonumber + ({ie\hbar }/{2Mc})\left[ {\mathbf B(R)}\times \nabla
_{R}-\nabla _{R}\times {\mathbf B(R)}\right] \cdot {\mathbf r}$
and $W_2 = ({e^{2}}/{8\mu c^{2}})B({\mathbf R})^{2}r^{2}$ are
terms related with the first and the second power in the magnetic
field strength, respectively. \cite{JAKF} In the above
Hamiltonian, $\mu = m_{e}^{\ast }m_{h}^{\ast }/ M $ is the exciton
reduced mass, $\xi = \left(m_{h}^{\ast }-m_{e}^{\ast }\right)/ M$,
and ${\mathbf L=} {\mathbf r\times } \left( -i\hbar \nabla
_{r}\right) $ is the exciton angular momentum associated to the
relative motion.

In the adiabatic approximation, the magnetic field in the spin
Hamiltonian [see $H^{m_z }({\mathbf r}_{e},{\mathbf r}_{h})$ in
Eq. (\ref{eq2})] can be expanded to zero order in the relative
coordinates, i.e., ${\mathbf B}({\mathbf r}_{e,h}) = {\mathbf
B}({\mathbf R})$. With this assumption, the exciton spin
Hamiltonian can be written as:

\begin{equation}
H^{m_z }({\mathbf R})=\mu _{B}\sum_{i=1}^{3}\left[
g_{e,i}S_{e,i}-\frac{1}{3} g_{h,i}J_{h,i}\right] B_{i}{\mathbf
(R)}-\frac{2}{3} \sum_{i=1}^{3}c_{i}S_{e,i}{}J_{h,i},
\end{equation}
where we introduced the following definitions for the exciton
heavy hole $g$ factor $g_{h,x}=3q_{x}$, $g_{h,y}=-3q_{y}$,
$g_{h,z}=6k_{z}+13.5q_{z}$. \cite{Kest}

Following the adiabatic approach, the total exciton wave function
can be written as $\Psi ^{m_z }({\mathbf R},{\mathbf
r},z_{e,h})=\Phi ({\mathbf r})\psi ({\mathbf R})F (z_{e,h}){\cal
L}^{m_z }({\mathbf R})$, where $\Phi ({\mathbf r})$ [$\psi
({\mathbf R})$] is the exciton wave function associated to the
relative (center-of-mass) motion, and ${\cal L}^{m_z}({\mathbf
R})$, $F (z_{e,h})$ is the exciton wave function corresponding to
the spin and to the confinement in the quantum well, respectively.
Please notice that in the case of the exciton motion in a
homogeneous magnetic field, the spin interaction with the magnetic
field can be solved separately from the center-of-mass and
relative motion coordinates. Therefore, we assume here that there
is no coupling between the wave function related to the exciton
spin contribution and those of the exciton center-of-mass and
relative motion. Finally, we obtain the following Schr\"{o}dinger
equations:

\begin{mathletters}
\label{alleq6}
\begin{eqnarray}
\left\{ H^{CM}({\mathbf R})+E^{r}(\gamma ,{\mathbf R},\nabla
_{R})+E^{m_z }({\mathbf R} ) + E^{\bot } - E\right\} \psi
({\mathbf R})=0, \label{hamcm} \\   \left\{ H_{\gamma
}^{r}({\mathbf r},{\mathbf R},\nabla _{R}) - E^{r}(\gamma
,{\mathbf R},\nabla _{R}) \right\} \Phi ({\mathbf r})=0,
\label{hamrel} \\ \left\{ H^{m_z }({\mathbf R}) - E^{m_z
}({\mathbf R}) \right\} {\cal L} ^{m_z }({\mathbf R})=0,
\label{hamspin}
\\ \left\{ H^{\bot }(z_{e,h}) -E^{\bot } \right\} F (z_{e,h})=0.
\label{hamwell}
\end{eqnarray}
\end{mathletters}
The variational parameter ($\gamma $) is chosen in such a way to
minimize the expectation value of $W(r,z_{e,h})$. The optimum
condition is found when

\begin{equation}
\triangle E^{'}=\left\langle \Phi ({\mathbf r}) F (z_{e,h})\left|
W(r,z_{e,h})\right| \Phi ({\mathbf r}) F (z_{e,h})\right\rangle =
0. \label{hamgamma}
\end{equation}
The above equation determines $\gamma _{\min }$ which is inserted
into the exciton relative motion energy $E^{r}(\gamma ,{\mathbf
R},\nabla _{R})$.
\subsection{The Nonhomogeneous Magnetic Field}

Here, we are interested in the possibility of exciton trapping by
using a confinement potential which is created by a circular
nonhomogeneous magnetic field. Experimentally, these magnetic
field profiles can, e.g., be created by the deposition of
nanostructured ferromagnetic disks (or superconducting disks) on
top of a semiconductor heterostructure with a homogeneous magnetic
field applied perpendicular to the $xy$ plane. This produces a
nonhomogeneous perpendicular magnetic field in the
plane.\cite{Dubonos,GeimPt} In this work, we consider a magnetized
disk on top of a GaAs/Al$_{x}$Ga$_{1-x}$As quantum well, which
creates a magnetic dipole type of profile. A sketch of the
experimental setup is shown in Fig. 1(a). This nonhomogeneous
magnetic field profile is given by the following
equation:\cite{JAKF}

\begin{eqnarray}
B_z(R) &=& B_{a} + 2B_0^D\frac{a(a+R)}{R\sqrt{(a+R)^2+d^2}}
\left\{-E\left(p^2\right)+ \left(1-\frac{p^2}{2}\right)K
\left(p^2\right)\right\} + B_0^D\frac{a\left( a^2-R^2+d^2
\right)}{R^2\sqrt{aR}}p^3 \nonumber \\ & \times & \left\{
-\frac{\partial }{\partial p^2}E(p^2)-\frac{1}{2}K(p^2) +
\left(1-\frac{p^2}{2}\right) \frac{\partial }{\partial
p^2}K(p^2)\right\}, \label{magfield}
\end{eqnarray}
where $ p=2\sqrt{a R} / \sqrt{\left(a+R\right)^2+d^2} $, $B_0^D= h
{\cal M} / a$ is the strength of the magnetization of the disk,
$B_{a}$ is the uniform applied field, $h$ is the disk thickness,
$a$ is the disk radius, $d$ the distance of the magnetic disk to
the quantum well, ${\cal M}$ the magnetization (in units of
$\mu_{0} {\cal M} / 4 \pi$, where $\mu_{0}$ is the permeability of
free space ), R the radial coordinate in the $xy$ plane, and
$K(x)$ $\left[E(x)\right]$ is the elliptic integral of first
(second) type. The magnetic field profile $B_z(R)$ is shown in
Fig. 1(b), for $a = 2$ $\mu$m, $d=0.2$ $\mu$m, $B_0^D = 0.05$ T,
and for different values of the external applied magnetic field
$B_{a}=0$, $-0.25$ T, and $0.35$ T. Notice that for this magnetic
field profile, the average magnetic field strength is zero, which
give us the additional possibility to apply a background field
$B_{a}$ to shift this magnetic field profile up and down. We will
show that this results into two different effective confinement
profiles for the exciton.

\section{Effective Mass and Confinement Potential}

In order to estimate the exciton confinement, we defined the
trapping energy ${E_{T}}$ as the energy difference between an
excitonic state in the homogeneous applied field $B_a$ and the
corresponding state in the nonhomogeneous magnetic field:
\begin{equation}
E_{T} = E^{r}(\gamma , B_{a}) + E^{m_{z}}(B_{a}) + E^{\bot } - E.
\end{equation}
Thus, the important terms (effective potentials) which determine
${E_{T}}$ in  Eq. (\ref{hamcm}) are only those appearing in Eqs.
(\ref{hamrel}-\ref{hamwell}) which are $B(R)$-dependent.
Therefore, the Hamiltonian corresponding to the exciton
confinement in the quantum well does not have to be solved in
order to obtain the trapping energy. Indeed the corresponding
result will cancel out in the $E_{T}$ calculation. All the well
confinement related contribution to the exciton trapping is then
determined through the $\gamma$ parameter [see Eq. (\ref{hamrel})
and Eq. (\ref{hamgamma})]. It is worthwhile to point out that if
we are interested in calculating the exciton binding energy, all
these terms [Eqs. (\ref{hamrel}-\ref{hamwell})] have to be
included.

Following Ref. [10], the exciton relative motion [see Eq.
(\ref{hamrel})] can be solved by perturbation techniques, where
all the $B(R)$-dependent terms are treated as perturbations. The
corresponding eigenvalues are:

\begin{eqnarray}
E^{r}(\gamma ,{ R},\nabla _{R})& =\frac{e\hbar }{2\mu c}\xi
m_{r}B_{z}({ R}) + \frac{e^{2}}{8\mu c^{2}} \beta_{m_{r}}^{n_{r}}
{\gamma^{-2}}B_{z}({ R})^{2} + \frac{e^{2}\mu}{ 2M^{2}c^{2}}
\alpha _{m_{r}}^{n_{r}} {\gamma^{-4}} \nabla _{R}\left[ B_{z}({
R})^{2}\nabla _{R} \right] , \label{enrrel}
\end{eqnarray}
with the following wave function:

\begin{equation}
\Phi _{0}^{n_{r},m_{r}}({\mathbf r})=A_{n_{r},m_{r}}\left(
\frac{2\gamma r}{a_{B}^{\ast} (n_{r} - 1/2)}\right) ^{|m_{r}|}
{\rm exp} \left( im_{r}\varphi -\frac{ \gamma r}{a_{B}^{\ast}
(n_{r} - 1/2) }\right) L_{n_{r}+|m_{r}|-1}^{2|m_{r}|}\left(
\frac{2\gamma r} {a_{B}^{\ast}(n_{r} - 1/2)}\right) ,
\end{equation}
where $n_{r}$, $m_{r}$ are quantum numbers of the exciton relative
motion, $a_{B}^{\ast }=\varepsilon \hbar ^{2}/\mu e^{2}$ is the
effective Bohr radius, $A_{n_{r},m_{r}} $ is the normalization
constant, $L_{n}^{(\alpha )}(x)$ is the generalized Laguerre
polynomial, and $\alpha _{m_{r}}^{n_{r}}$ and $ \beta
_{m_{r}}^{n_{r}}$ (in units of ${a_{B}^{\ast}}^4$ and
${a_{B}^{\ast}}^2$, respectively) are constants related to the
relative quantum numbers $n_{r}$, $m_{r}$.\cite{JAKF} In writing
the above energy for the relative motion [Eq. (\ref{enrrel})] we
have neglected the field independent term, $ -\gamma^2 R_y^* /
(n_r -1/2)^{2} $ (where $R_{y}^{\ast }=\mu e^{4}/2\varepsilon
^{2}\hbar ^{2}$ is the effective Rydberg), which does not give any
contribution to the exciton trapping energy. In solving the
exciton relative motion we assumed that the magnetic field
intensity is such that we are in the weak-field regime, i.e.,
$\hbar \omega^{*}_{c} < 2 R^{*}_{y}$ where $\omega^{*}_{c} = e B /
\mu c$ is the cyclotron-resonance frequency. For GaAs this implies
$B < 5 $ T.

For the sake of simplicity, instead of solving Eq.
(\ref{hamgamma}) to obtain $\gamma$, we used the results of
St\'{e}b\'{e} and A. Moradi\cite{Stebe} and of Andreani and
Pasquarello\cite{Lucio} to determine the zero magnetic field
binding energy of excitons in GaAs/Al$_{0.3}$Ga$_{0.7}$As quantum
wells.\cite{gamma} We found that the gamma parameter only slightly
influences the trapping energy and this as a coefficient in two
smaller terms (diamagnetic and mass correction) in the exciton
relative motion energy [see Eq. (\ref{enrrel})]. For a complete
description of the dependence of the $\gamma$ parameter with the
quantum well width and also with an applied homogeneous magnetic
field (which is negligible in the weak field regime) we refer to
Ref. [19].

The eigenenergy of the spin Hamiltonian for a magnetic field
parallel to the z direction can be straightforwardly calculated
from Eq. (\ref{hamspin}):

\begin{equation}
E^{m_z }({ R}) = \pm \frac{1}{2}\sqrt{\mu _{B}^{2}\left[ \left(
-1\right)^{m_z +1}g_{e,z}+g_{h,z}\right]^{2}B_{z}( {R})^{2}+\left[
c_{x}-\left( -1\right)^{m_z +1}c_{y}\right]^{2}}, \label{enspin}
\end{equation}

In the above energy, we neglected the $B(R)$-independent term,
i.e., the e-h exchange energy in the z-direction $\left(
-1\right)^{m_z +1}c_{z}/2 $,\cite{Kest} which does not contribute
to the calculus of $E_{T}$. The spin wave function is a linear
combination [$|{\cal L}^{m_z}({\mathbf R}) \rangle = (|+\rangle
\pm |-\rangle) / \sqrt{2}$] of the exciton spin states $m_z$:

\begin{equation}
|{\cal L}^{m_z}({ R}) \rangle = \frac{|\left| m_z \right| \rangle
+\left(Q \pm \sqrt{1+Q^{2}}\right) |-\left| m_z \right| \rangle
}{\sqrt{2 \left( 1 + Q^{2} \pm Q\sqrt{1+Q^{2}}\right) }} ,
\end{equation}
where $Q = \mu _{B}\left[ \left( -1\right)^{m_z
+1}g_{e,z}+g_{h,z}\right] B_{z}({R}) {\Big /} \left[ c_{x}-\left(
-1\right)^{m_z +1}c_{y}\right]$.

We can now use all the results of the exciton relative motion,
spin interaction and well confinement [Eqs. (\ref{enrrel}),
(\ref{enspin}), and (\ref{hamgamma}), respectively] to solve the
exciton center-of-mass equation [see Eq. (\ref{hamcm})]. Now, the
eigenenergies $E^{r}(\gamma ,{R},\nabla _{R})$ and $E^{m_z }(
{R})$ act like an effective potential and an effective mass in the
center-of-mass equation of motion:

\begin{eqnarray}
\Biggl\{-\frac{\hbar ^{2}}{2M}\nabla _{R}\left\{
1-\frac{e^{2}\mu}{\hbar ^{2} M c^{2}} \alpha_{m_{r}}^{n_{r}}
{\gamma^{-4}}B_{z}({R} )^{2}\right\} \nabla _{R}+\frac{e^{2}}{8\mu
c^{2}} \beta_{m_{r}}^{n_{r}} \gamma^{-2}
B_{z}({R})^{2}+\frac{e\hbar }{2\mu c}\xi B_{z}({R} )m_{r}
\nonumber \\ \pm \frac{1}{2} \sqrt{\mu_{B}^{2}\left[ \left(
-1\right)^{m_z +1} g_{e,z}+g_{h,z}\right]^{2}B_{z}(
{R})^{2}+\left[ c_{x}-\left( -1\right)^{m_z
+1}c_{y}\right]^{2}}-E\Biggl\}\psi ({R}) =0. \label{HamCM}
\end{eqnarray}

Our magnetic field is parallel to the z-direction and has $\varphi
$-symmetry, i.e., $B_{z}({\mathbf R})=B_{z}(R)$ [see Eq.
(\ref{magfield})]. Then, we can use the cylindrical symmetry of
our problem to write the exciton center-of-mass wave function as
$\psi ({\mathbf R})=e^{-im_{R}\varphi }\psi (R)$, where $m_{R}$ is
the quantum number for the angular momentum of the exciton
center-of-mass motion. Inserting the above wave function in Eq.
(\ref{HamCM}) and writing the $\nabla $ operator in cylindrical
coordinates, Eq. (\ref{HamCM}) can be written as follows:

\begin{equation}
\left\{ -\frac{\hbar ^{2}}{2}\frac{1}{R}\frac{d}{dR}\left[
\frac{R}{ M^{eff}(R)}\frac{d}{dR}\right] +V^{eff}(R)-E\right\}
\psi (R)=0, \label{end}
\end{equation}
where

\begin{equation}
M^{eff}(R)=\frac{M}{1-\frac{e^{2}\mu}{ \hbar^{2} M c^{2}}
\alpha_{m_{r}}^{n_{r}} {\gamma^{-4}}B_{z}(R)^{2}}, \eqnum{15a}
\label{masseff}
\end{equation}
is the exciton effective mass, and

\begin{eqnarray}
V^{eff}(R) &=& \frac{\hbar ^{2}}{2}\frac{1}{M^{eff}(R)}
\frac{m_{R}^{2}}{R^{2}} + \frac{e^{2}}{8\mu c^{2}}
\beta_{m_{r}}^{n_{r}} \gamma^{-2} B_{z}(R)^{2} + \frac{e\hbar
}{2\mu c}\xi m_{r}B_{z}(R) \nonumber
\\  &\pm& {\frac{ 1}{2}\sqrt{\mu _{B}^{2}\left[ \left(
-1\right)^{m_z+1}g_{e,z} + g_{h,z}\right]^{2} B_{z}(R)^{2}+\left[
c_{x}-\left( -1\right) ^{m_z +1}c_{y}\right]^{2}} } , \eqnum{15b}
\label{Poteff}
\end{eqnarray}
is the effective confinement potential of the exciton
center-of-mass motion. In the above equation, the different terms
correspond to the centrifugal, diamagnetic, orbital momentum, and
spin contribution, respectively. The centrifugal term is related
with the exciton effective mass and the angular momentum of the
center-of-mass motion, which we found to give the smallest
contribution to the effective potential. The magnetic field
squared dependence is present in the effective mass and in the
diamagnetic term of the effective potential. Notice that all the
direct contributions of the quantum well confinement is included
through the $\gamma$ parameter, which only gives a small increase
in the magnetic field squared intensity. As occurs in the trapping
of atoms, the two more important terms are the angular momentum
and spin contributions. The first is only important for the
excited states. The sign in the spin contribution is related to
the spin quantum state $m_z = \pm 1, \pm 2$, which characterizes
the spin Zeeman splitting.
\section{Numerical Results and Discussion}

We have calculated the trapping energy and wave function of
excitons in a GaAs/Al$_{0.3}$Ga$_{0.7}$As quantum well in a
nonhomogeneous dipole type of magnetic field. The numerical
solution of Eq. (\ref{end}) was obtained by using a discretization
technique. We use electron and heavy-hole $g$ factors which depend
on the quantum well width $L$, but we neglect its magnetic field
dependence since only small magnetic fields are
considered.\cite{Snel92,Snel94,Snel96} We take advantage of the
system symmetry to assume that $c_{x}=-c_{y}$. \cite{Snel94} The
electron and heavy hole mass, and the dieletric constant used in
this work are the same as in Ref. [23,24] ($ m_{e}^{*}/m_{0} =
0.067 $, $ m_{h}^{*}/m_{0} = 0.34 $, and $ \varepsilon = 12.5 $).
The confinement energy level of the exciton in the quantum well
does not directly enter into the trapping energy, but it enters
indirectly through the $\gamma$ parameter which depends on the
confinement state of the exciton and on its relative motion state.
As an example we took a ferromagnetic disk with radius $a=2$
$\mu$m, which is placed a distance $d=0.2$ $\mu$m above the
quantum well.

The effective potential and effective mass for the exciton ground
state is shown in Fig. \ref{figure2} as a function of the radial
coordinate $R$, for a quantum well width of $L = 90$ \AA, for the
situation in which an uniform applied field of (a) $B_{a}=0.35$ T
and (b) $B_{a}=-0.25$ T is presented, where the strength of the
disk magnetization is $B_{0}^{D}=0.05$ T.  The Zeeman effect can
be seen in the effective potential by the shift in the
corresponding spin states $m_{z}= \pm 1$ (dashed and dotted
curves), as compared to the spinless state (solid curves). Notice
that there is an interesting change of both sign and curvature in
the effective potential associated to the spin states $m_{z}= \pm
1$, which can be observed by comparing the dotted curves ($m_{z}
=+1$) with the dashed curves ($m_{z}=-1$) depicted in Fig.
\ref{figure2}. This occurs due to the fact that the diamagnetic
contribution to the effective potential [see Eq. (\ref{Poteff})]
is usually smaller than the Zeeman contribution. Then, the latter
dictates the behavior of the effective confinement potential,
changing the confinement of the trapped exciton from a centered
structure to a ring-like structure.

The quantum well confinement effects on the exciton motion are
very important not because of the confinement itself, but due to
the large dependence of the exciton $g$ factor on the quantum well
width [see inset of Fig. \ref{figure3}(a)]. The exciton trapping
energy dependence on the strength of the magnetization of the disk
($B_0^D$), for a homogeneous applied field of $B_a = -0.25$ T, for
quantum wells widths of $L = 50$ {\AA} and $L = 100$ {\AA} is
shown in Fig. \ref{figure3}(a) and Fig. \ref{figure3}(b),
respectively. In both figures we showed the results for the case
when the exciton center-of-mass quantum number is $n_R = 1$ with
$m_R = 0$ (curves) and $m_R = 1$ (symbols), and when the relative
quantum numbers are ($n_r = 1$, $m_r = 0$), for the ground state
level of the exciton well confinement, and for spin quantum number
$m_{z} = 0, \pm 1 $ (i.e., $\sigma^{\pm}$ polarized states), $\pm
2$ (i.e., the dark excitons). The results for the dark exciton are
only shown for reference. They are not optically active and they
will not be further considered. Notice that the exciton $g$ factor
increases the trapping energy as much as by a factor of 20 and
that the behavior of the Zeeman splitting is strongly influenced
by the nonhomogeneous magnetic field [compare the dashed and
dotted curves in Fig. \ref{figure3}(a) and \ref{figure3}(b)]. Also
notice that for large $B_{0}^{D}$, the trapping energy starts to
decrease because the nonhomogeneous field created by $B_{0}^{D}$
can now be comparable to the homogeneous applied field $B_{a}$,
which decreases the confinement region in the effective potential
[see Fig. \ref{figure2}(b) and Fig. 1(b)]. The energy of the $m_R
= 1$ states [the centrifugal term in the effective potential of
Eq. (\ref{Poteff})] are only slightly different from the energy of
the correspondent non-excited state.

The dependence of the exciton ground state trapping energy on the
quantum well width and on the magnetization strength $B_{0}^{D}$
is shown in Figs. \ref{figure4}, \ref{figure5}(a), and
\ref{figure5}(b), for the spin quantum numbers $m_{z} = 0$, $-1$,
and $+1$, respectively, and for an applied field $B_{a} = 0.35$ T.
For thin wells (widths smaller than approximately 60 {\AA}) the
exciton trapping energy is strongly sensitive to the
nonhomogeneous magnetic field and to the size of the well. As the
well width increases, the trapping energy becomes basically
independent of the well confinement, and the magnetic field is
only a small perturbation [see Figs. \ref{figure5}(a) and
\ref{figure5}(b)]. There is a minimum in the exciton trapping
energy when the quantum well width is near 100 {\AA} which is due
to the fact that the $g$ factor is approximately equal to zero
[see inset of Fig. \ref{figure3}(a)]. The spin Zeeman splitting as
due to the nonhomogeneous magnetic field does not follow the same
behavior of shifting up and down the energy as occurs in the
homogeneous field case [compare Figs. \ref{figure4},
\ref{figure5}(a) and \ref{figure5}(b)]. The $\sigma^{-}$ polarized
state [Fig. \ref{figure5}(a)] is always shifted up as compared to
the spinless situation (Fig. \ref{figure4}), but the $\sigma^{+}$
polarized state can be shifted down, but also up [see Fig.
\ref{figure5}(b) or dotted curves in Fig. \ref{figure3}(a) and
\ref{figure3}(b)], depending on the relation between the quantum
well width (exciton $g$ factor) and the magnetization strength
$B_{0}^{D}$. This is due to the change in the confinement region
of the effective potential related with the $m_z = +1$ spin state,
as previously described in our discussion of Fig. \ref{figure2}.
It is quite remarkable that the exciton spin interaction with the
nonhomogeneous magnetic field can be responsible for increases in
the exciton trapping energy as large as a factor of 100, as
compared to the spinless situation of Fig. \ref{figure4}. It is
important to highlight that the considered magnetic fields and
quantum well widths are in the range which are currently available
experimentally.

The trapping energy of the exciton excited states as a function of
the strength of the disk magnetization $B_{0}^{D}$ are shown in
Fig. \ref{figure6} for ($n_{R} = 1$, $m_{R} = 0$) with relative
quantum number $n_r = 2$, for the first excited state of the
exciton well confinement, for a quantum well width $L = 50$ {\AA},
for $B_{a} = -0.25$ T, with the following relative angular quantum
numbers: (a) $m_r = -1$ , (b) $m_r = 0$, and (c) $m_r = +1$, and
for $B_{a} = -0.25$ T, for (d) $m_r = -1$ , (e) $m_r = 0$, and (f)
$m_r = +1$. The interaction of the exciton angular and spin
momentum is responsible for several interesting effects, mainly in
the case of negative $B_{a}$ where the Zeeman splitting exhibits
different behavior for each angular quantum number of the relative
motion. The trapping energy for a positive applied field always
increases linearly with increasing nonhomogeneous magnetic field
intensity $B_{0}^{D}$, which is not the case for the results with
negative $B_{a}$. The linear increase is due to the fact that the
angular momentum term is the dominant term in the exciton
Hamiltonian, and it has a linear dependence on the magnetic field.
The decrease of the trapping energies in Figs. \ref{figure6}(a)
and \ref{figure6}(b) can be explained by the competition between
$B_{a}$ and $B_{0}^{D}$, as discussed previously in connection
with Fig. \ref{figure3}.

The contour plot of the conditional probability to find the
electron somewhere in the $xy$ plane for the case of the exciton
ground state $|\psi ({\mathbf R}) \Phi ({\mathbf r})|^2$, in the
presence of an applied homogeneous magnetic field $B_{a} = -0.25$
T, is shown in Fig. \ref{figure7} for the spin quantum number
$m_{z} = -1$ and in Fig. \ref{figure8} for $m_{z} = +1$. In both
pictures, we considered a quantum well width $L = 90 $ {\AA} and
the strength of the disk magnetization was $B_{0}^{D} = 0.05$ T.
The hole (indicated by the symbol in the figures with a cross in
the middle) is fixed in the position ${\mathbf r}_{h} = (0,0)$ in
Figs. \ref{figure7}(a) and \ref{figure8}(a), and in ${\mathbf
r}_{h} = (0.5,0)$ $\mu$m in Figs. \ref{figure7}(b) and
\ref{figure8}(b). It can clearly be seen that the spin orientation
changes the exciton confinement from a centered structure to a
ring like structure (compare Figs. \ref{figure7} and
\ref{figure8}). Also notice that the electron tries to follow the
hole but this is partially worked against by the attraction to the
minimum of the effective potential (see Fig. \ref{figure2}) in the
case of a centered wave function (see Fig. \ref{figure7}). In the
ring-like structure, the electron moves to the hole nearest
position (see Fig. \ref{figure8}).
\section{Conclusions}
The possibility of carrier trapping in a well-defined confinement
region in a semiconductor heterostructure open the possibility to
numerous studies for electronic phase transitions, confined
exciton gases and excitonic molecules.\cite{Butov,Negoit} The
trapping of excitons in semiconductor systems of reduced
dimensionality have also been proposed as a method of observing
Bose-Einstein condensation of excitons.\cite{Butov2} The
experimental efforts to observe exciton condensation in
semiconductors were concentrated on the analysis of the PL line
shape and the transport of excitons. Previous PL experiments (see,
e.g., Ref. [11] and references therein) of the exciton trapping by
using nonhomogeneous magnetic fields have shown that excitons are
driven to regions of minimum field, but they could not detect the
direct effect of the field gradient on the exciton motion. They
concluded that the experimental conditions had to be improved by
taking better samples which have larger exciton mobility, and a
larger nonhomogeneous magnetic field strength was necessary.

We have investigated the exciton trapping in a quantum well in the
presence of a nonhomogeneous dipole type of magnetic field. The
effects of the well confinement and of the exciton interaction
with the nonhomogeneous magnetic field taking into account spin
states were analyzed and discussed. As compared to our previous
results in which the spin of the exciton was not taken into
account, we found a substantial increase of the trapping energy of
the excitons by using nonhomogeneous magnetic fields. We also
showed that the obtained energies, as well as the considered
magnetic field strength and well widths are currently experimental
obtainable.

Our results show that the trapping energy is strongly dependent on
the quantum well width and shape of the nonhomogeneous magnetic
field profile, but that a increase in the magnetic field intensity
is not always related with a stronger confinement (as occurs in
the uniform magnetic field situation), as previously discussed in
Sec. IV. Our main result is that the spin of the exciton is
responsible for an increases in the trapping energy, which can be
as large as a factor of 100. This can be used to increase the
experimental conditions by choosing a suitable set of parameters
(quantum well width and magnetic field profile) in order to
maximize the exciton trapping. Furthermore, the spin interaction
with the magnetic dipole type of profile can change the exciton
spatial localization of a centered structure to a ring-like
structure, increasing the energy level corresponding to the $m_{z}
= +1$ spin state, which suggests that the two lines of the
$\sigma^{\pm}$ polarized states will be very close in energy in
the PL spectra, which may make it difficult for its
identification. For the case of narrow wells, exciton localization
due to quantum well width fluctuations will also be present. This
effect was not considered in the present paper but should be
easily distinguished experimentally with the present exciton
trapping by nonhomogeneous magnetic field, by changing the
strength of the magnetization of the magnetized disk.

Our results open up a new path for experiments on the confinement
of excitons, which should reveal new kinds of exciton trapping and
increase the knowledge about the exciton interaction with
nonhomogeneous magnetic fields.
\section{Acknowledgments}
This research was supported by the Flemish Science Foundation
(FWO-VI), the IUAP (Belgium), the "Onderzoeksraad van de
Universiteit Antwerpen", and by the Inter-university
Micro-Electronics Center (IMEC, Leuven). J. A. K. Freire was
supported by the Brazilian Ministry of Culture and Education
(MEC-CAPES) and F. M. Peeters was supported by the FWO-Vl. V. N.
Freire and G. A. Farias would like to acknowledge the partial
financial support received from CNPq, the Funding Agency of the
Cear\'a State in Brazil (FUNCAP), and the Brazilian Ministry of
Planning through FINEP.
\begin{figure}
\caption{(a) Experimental setup to create a magnetic dipole type
of profile in the $xy$ plane of the exciton motion in a quantum
well. Here $a$ ($h$) is the disk radius (thickness), $d$ is the
distance of the disk to the quantum well, ${\cal M}$ the
magnetization, $B_a$ is an uniform applied field, and $B_0^D$ is
the strength of the disk magnetization. (b) The magnetic dipole
type of profile as a function of the radial coordinate R for
different values of the background magnetic field.}
\label{figure1}
\end{figure}
\begin{figure}
\caption{Effective potential, V$^{eff}$(R), for the exciton ground
state, for the spinless situation $m_{R} = 0$ (solid), and
respective effective mass, M$^{eff}$(R), (dashed-dotted), and
V$^{eff}$(R) for spin quantum states $m_z = +1$ (dotted), and $m_z
= -1$ (dashed) of the magnetized disk, as a function of the radial
coordinate R, for a homogeneous applied magnetic field of (a) $B_a
= 0.35$ T and (b) $B_a = -0.25$ T, for $z/a$ = 0.1 and $a = 2$
$\mu$m, for a strength of the disk magnetization of $B_0^D = 0.05$
T, and a quantum well width $L = 90$ {\AA}.} \label{figure2}
\end{figure}

\begin{figure}
\caption{Exciton trapping energy, $E_{T}$, for $n_R = 1$,
$(n_r,m_r) = (1,0)$, as a function of the magnetization strength
$B_{0}^{D}$, in the presence of an applied field $B_a = -0.25$ T
and a quantum well width of (a) $L = 50$ {\AA} and (b) $L = 100$
{\AA}, with $m_R = 0$ (curves) and $m_R = 1$ (symbols), and for
spin quantum numbers $m_z = 0$ (solid curves and triangles down),
$m_z = +1$ (dotted curves and squares), $m_z = -1$ (dashed curves
and triangles up), $m_z = +2$ (short-dash curves and stars), and
$m_z = -2$ (dashed-dotted curves and circles). In the inset of (a)
the dependence of the $g$ factor on the quantum well width is
shown.} \label{figure3}
\end{figure}

\begin{figure}
\caption{Exciton trapping energy, $E_T$, for the exciton ground
state as a function of the strength of the disk magnetization
$B_{0}^{D}$ and the quantum well width $L$, for the spinless
exciton ($m_z = 0$) and for an applied background field of $B_a =
0.35$ T.} \label{figure4}
\end{figure}

\begin{figure}
\caption{The same as in Fig. 4 but now for spin quantum numbers
(a) $m_z = -1$ and (b) $m_z = +1$.} \label{figure5}
\end{figure}

\begin{figure}
\caption{Exciton trapping energy, $E_T$, for $(n_R,m_R) = (1,0)$,
$m_z = 0$ (solid curves), $m_z = +1$ (dotted curves), $m_z = -1$
(dashed curves) as a function of $B_{0}^{D}$ in the presence of an
applied field $B_a = -0.25$ T for relative quantum numbers (a)
$(n_{r},m_{r}) = (2,-1)$, (b) $(n_{r},m_{r}) = (2,0)$, and (c)
$(n_{r},m_{r}) = (2,+1)$, and for $B_a = 0.35$ T for $n_{r} = 2$
and (d) $m_{r} = -1$, (e) $m_{r} = 0$, and (f) $m_{r} = +1$. The
quantum well width is $L = 50$ {\AA}.} \label{figure6}
\end{figure}

\begin{figure}
\caption{Contour map of the electron conditional probability
$|\psi ({\mathbf R}) \Phi ({\mathbf r})|^2$ in the exciton ground
state, for spin quantum number $m_z = -1$, quantum well width $L =
90$ {\AA}, magnetization strength $B_{0}^{D} = 0.05$ T, and
uniform applied field $B_a = -0.25$ T. We fixed the hole in
position (a) ${\mathbf r}_h = (0,0)$ and (b) ${\mathbf r}_h =
(0.5,0)$ $\mu$m.} \label{figure7}
\end{figure}

\begin{figure}
\caption{The same as in Fig 7 but now for spin quantum number $m_z
= +1$.} \label{figure8}
\end{figure}


\begin{references}
\bibitem{gamma} For the calculus of $\gamma$, we compare the energy
of the zero field term in the exciton relative motion [$ -\gamma^2
R_y^* / (n_r -1/2)^{2} $] with the corresponding data in the
literature for the exciton binding energy, i.e., $\gamma^2 =
E_{b}^{lit.} / [ R_y^* / (n_r -1/2)^{2} ]$.
\end{references}
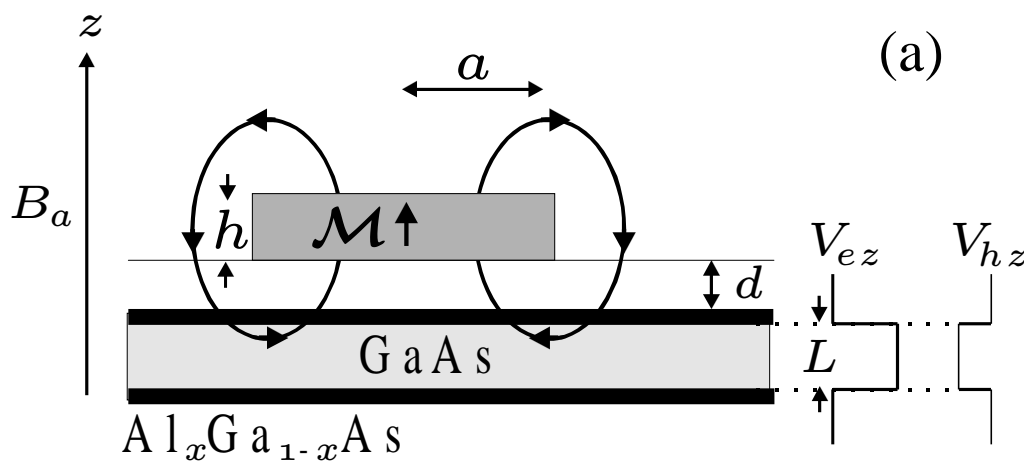
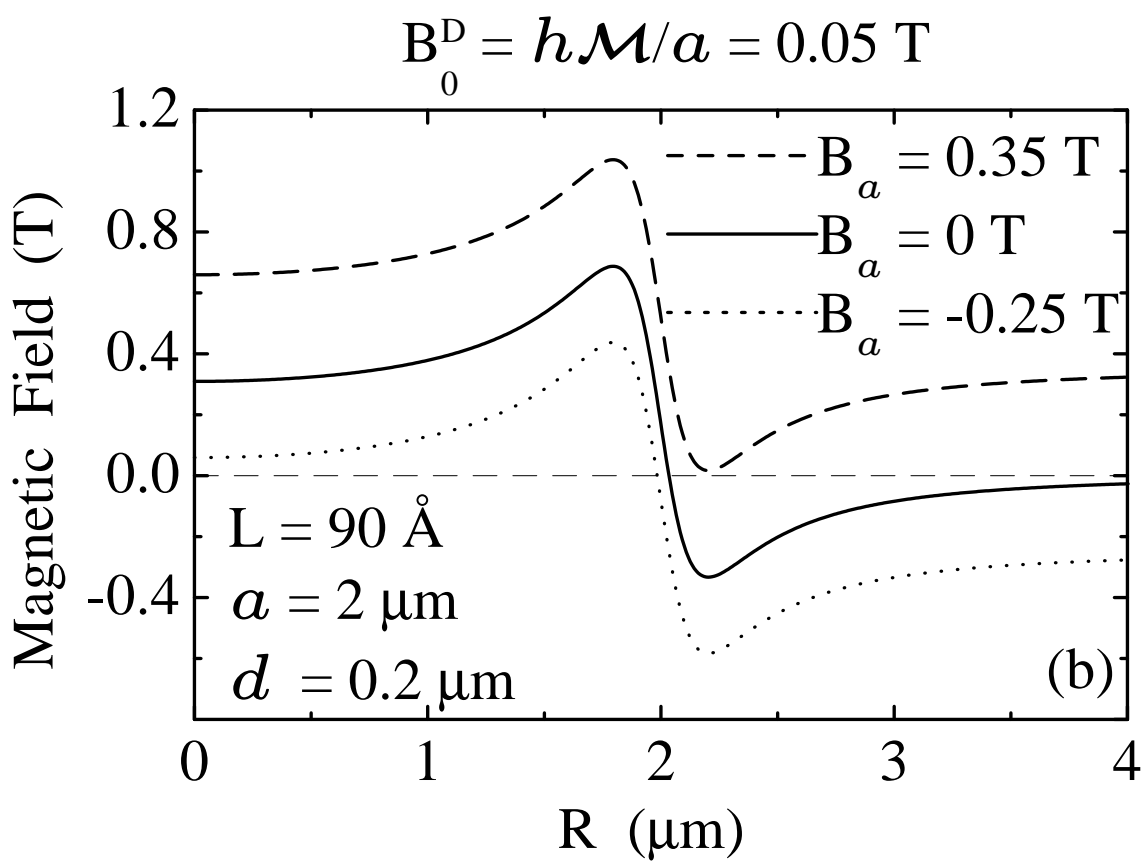
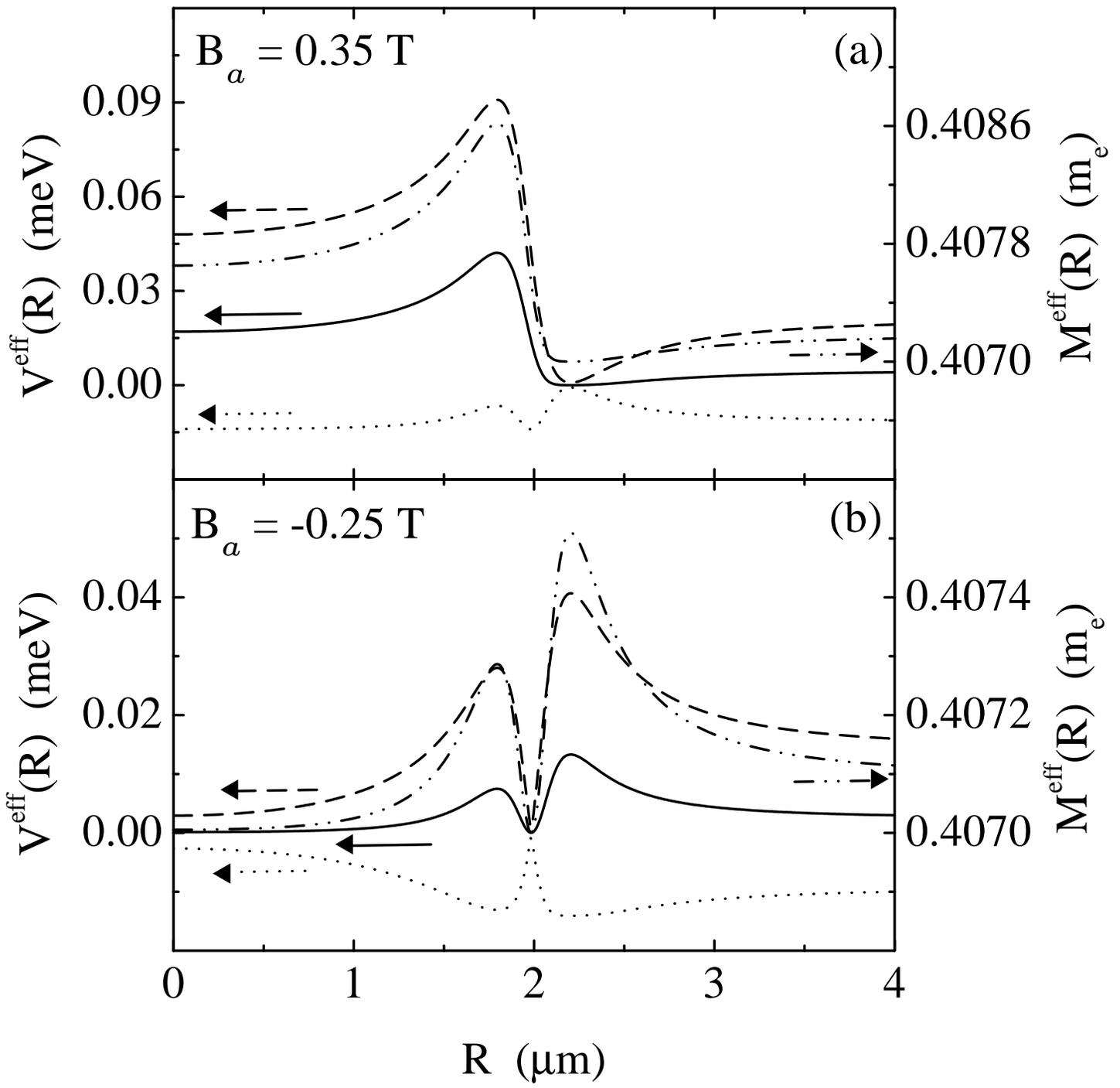
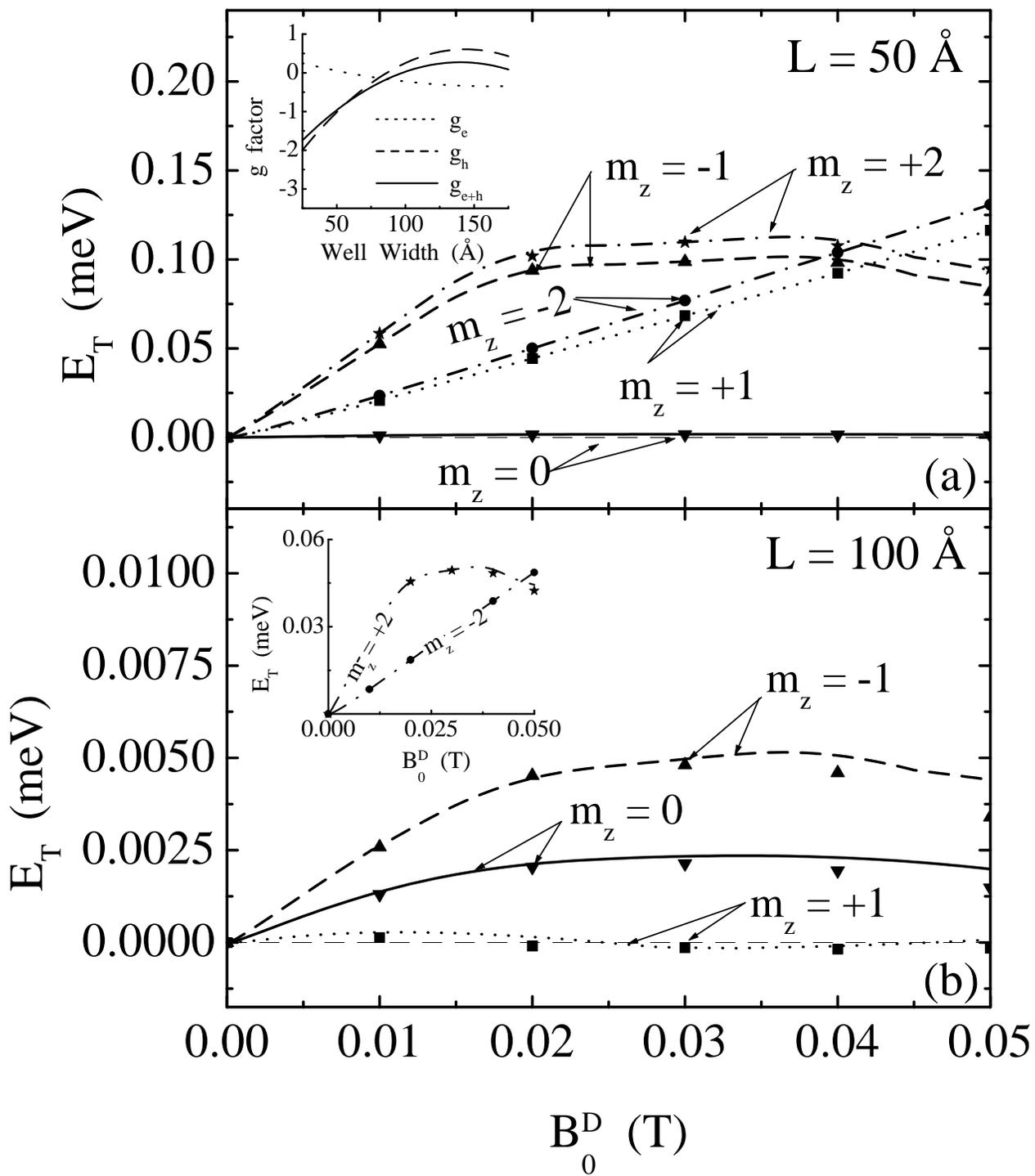
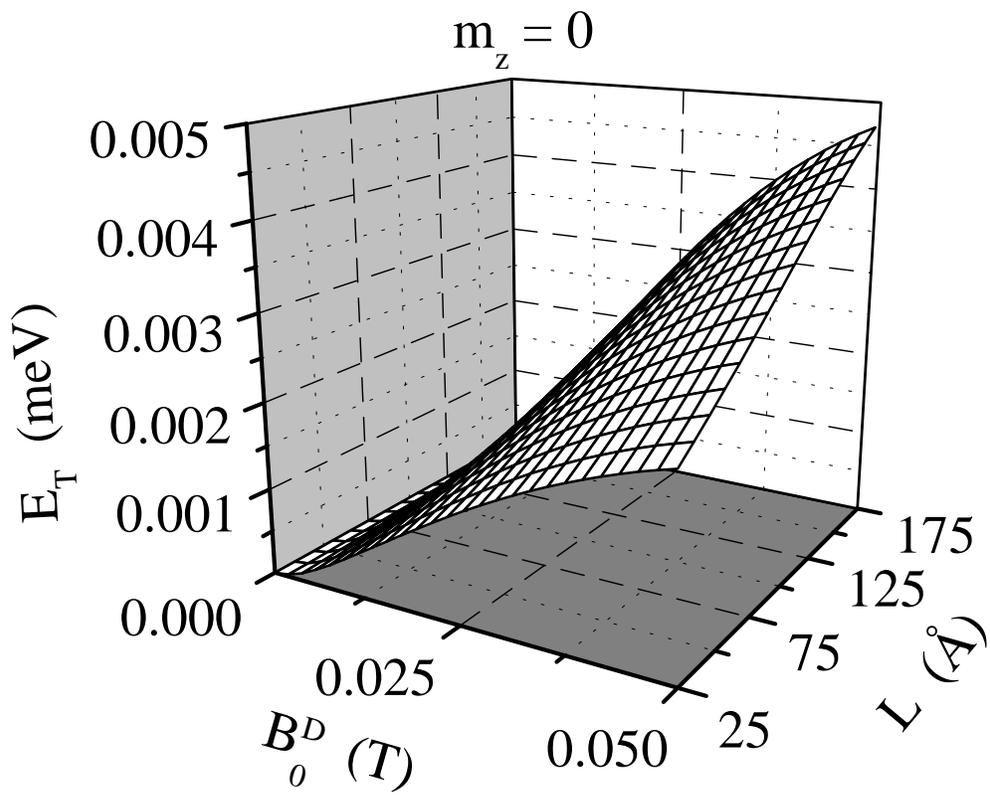
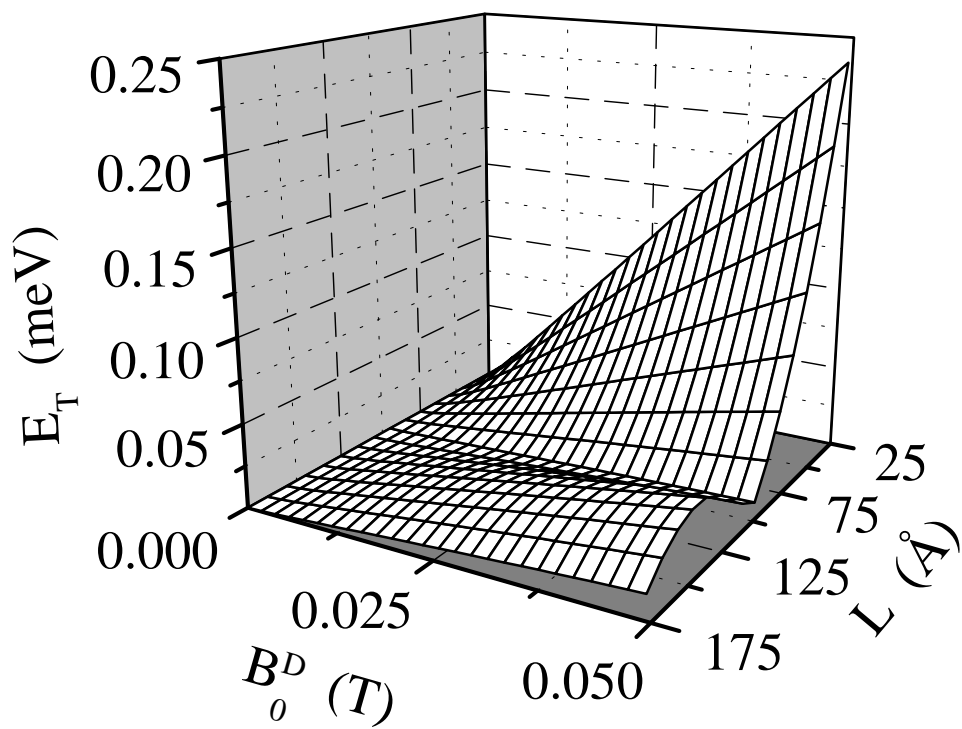
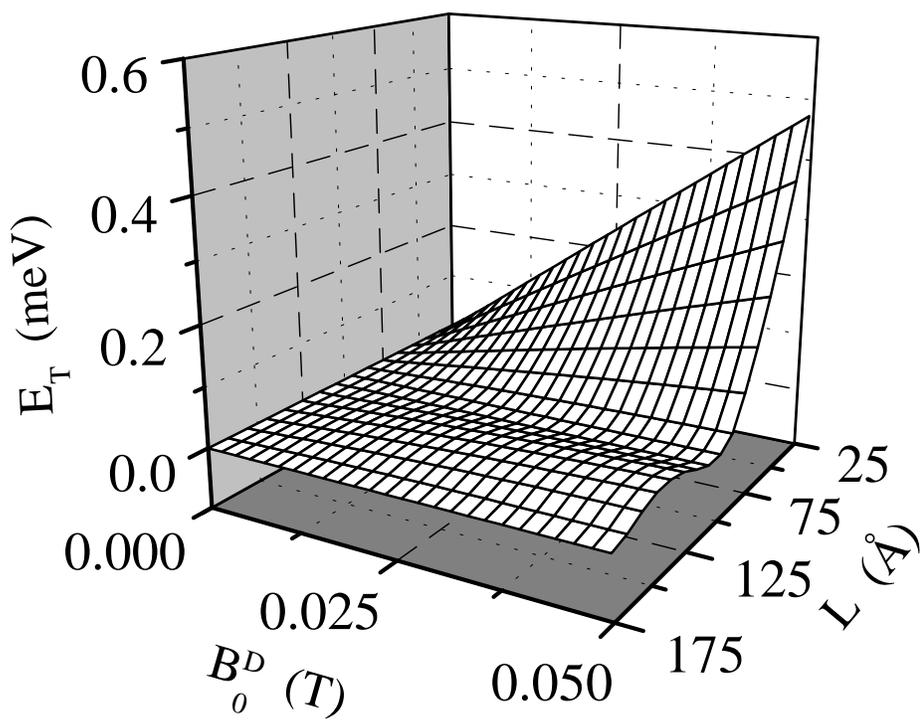
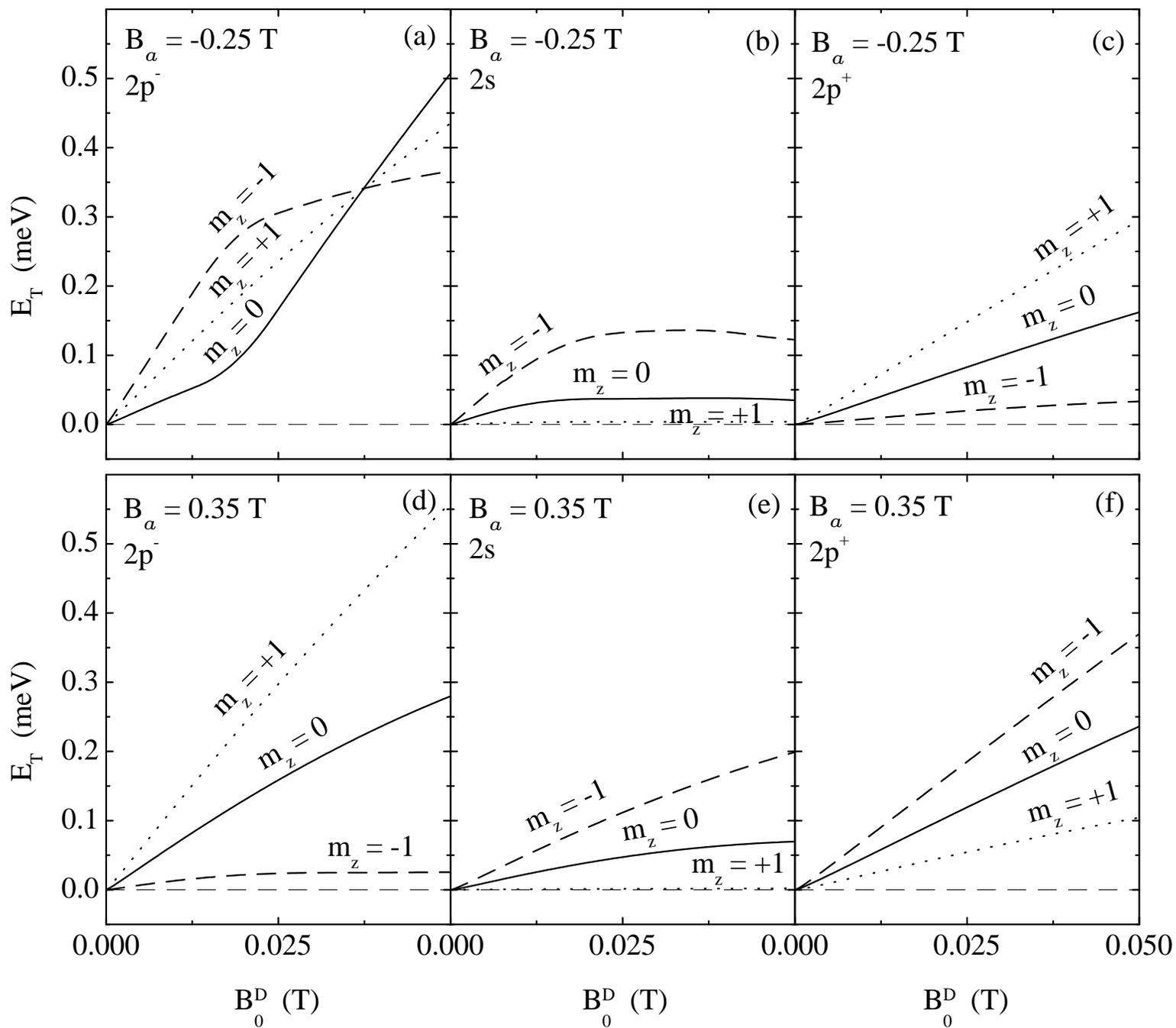
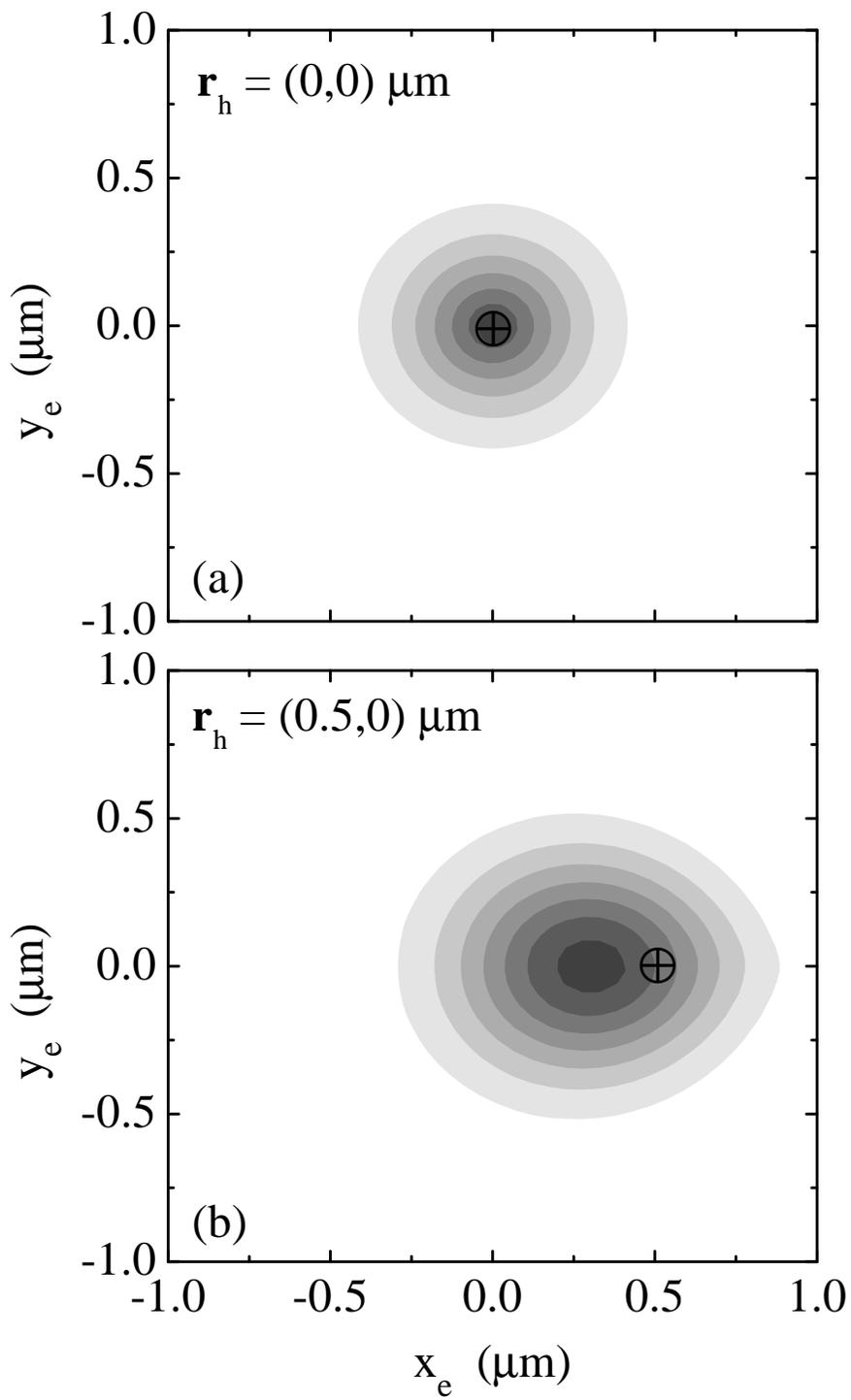
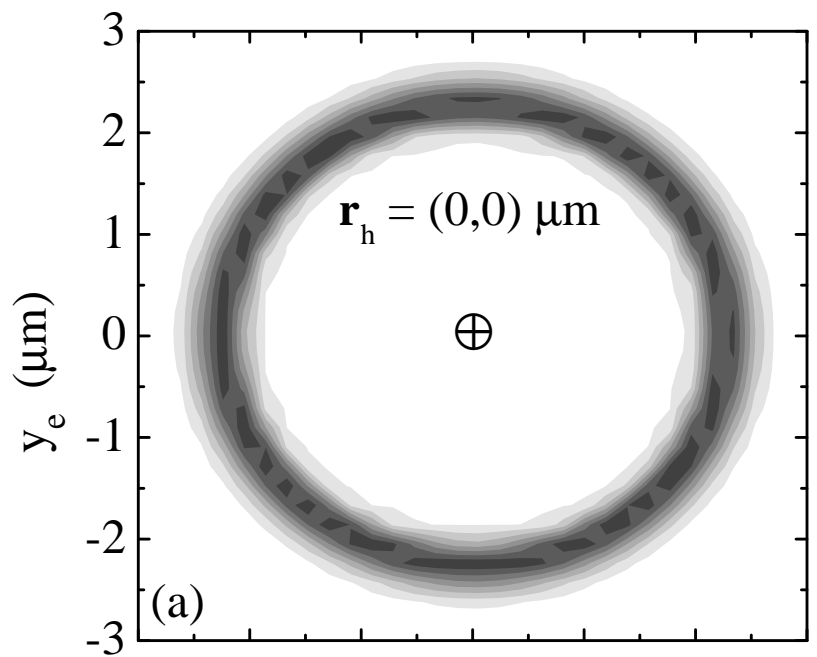
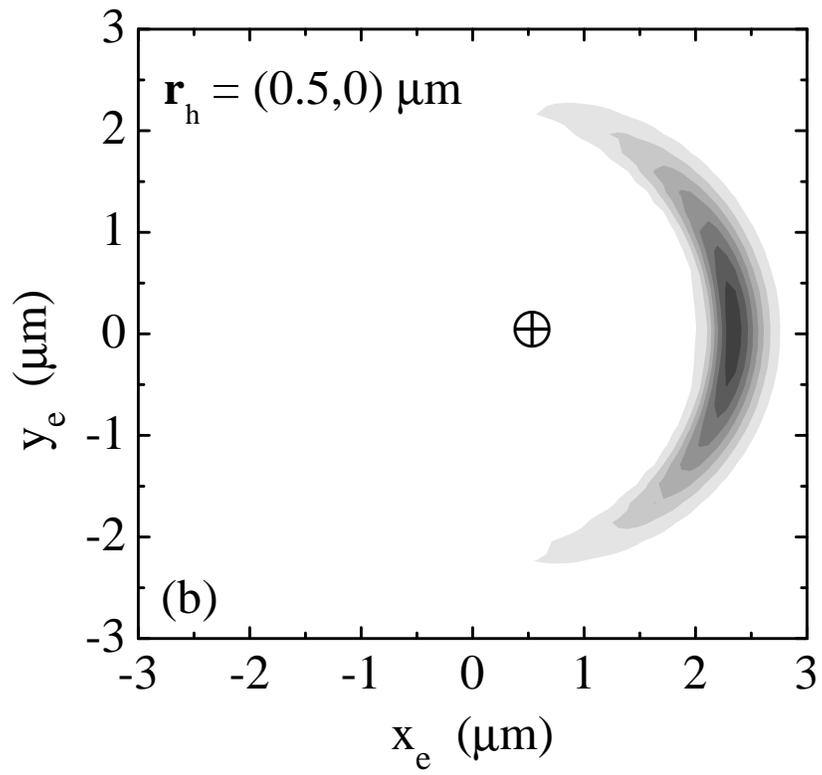
\end{document}